\documentclass[reprint,superscriptaddress,amsmath,amssymb,aps,prl]{revtex4-2}
\usepackage{graphicx}
\usepackage{dcolumn}
\usepackage{bm}
\usepackage{tabularx}
\usepackage{amsmath}
\usepackage{xcolor}
\usepackage{hyperref}

\hypersetup{
    colorlinks=true,
    linkcolor=blue,
    filecolor=blue,
    urlcolor=blue,
    citecolor=blue
}

\emergencystretch=\maxdimen
\begin{document}

\title{Thermoelectricity evidence for quantum criticality in clean infinite-layer nickelate films}

\author{Xu Zhang}
  \affiliation{State Key Laboratory of Surface Physics, Department of Physics, Fudan University, Shanghai 200438, China}
  \affiliation{Shanghai Research Center for Quantum Sciences, Shanghai 201315, China}
  
\author{Chihao Li}
  \affiliation{State Key Laboratory of Surface Physics, Department of Physics, Fudan University, Shanghai 200438, China}

\author{Mingwei Yang}
  \affiliation{Department of Physics, City University of Hong Kong, Kowloon, Hong Kong SAR, China}
  \affiliation{Shenzhen Research Institute of City University of Hong Kong, Shenzhen 518057, China}
  
\author{Yan Zhao}
  \affiliation{School of Physics, University of Electronic Science and Technology of China, Chengdu, China}

\author{Zhitong An}
  \affiliation{State Key Laboratory of Surface Physics, Department of Physics, Fudan University, Shanghai 200438, China}

\author{Danfeng Li}
  \email{Contact author: danfeng.li@cityu.edu.hk}
  \affiliation{Department of Physics, City University of Hong Kong, Kowloon, Hong Kong SAR, China}
  \affiliation{Shenzhen Research Institute of City University of Hong Kong, Shenzhen 518057, China}

\author{Liang Qiao}
  \email{Contact author: liang.qiao@uestc.edu.cn}
  \affiliation{School of Physics, University of Electronic Science and Technology of China, Chengdu, China}

\author{Haichao Xu}
  \affiliation{State Key Laboratory of Surface Physics, Department of Physics, Fudan University, Shanghai 200438, China}
  \affiliation{Shanghai Research Center for Quantum Sciences, Shanghai 201315, China}
  
\author{Rui Peng}
  \email{Contact author: pengrui@fudan.edu.cn}
  \affiliation{State Key Laboratory of Surface Physics, Department of Physics, Fudan University, Shanghai 200438, China}
  \affiliation{Shanghai Research Center for Quantum Sciences, Shanghai 201315, China}
  
\author{Donglai Feng}
  \affiliation{New Cornerstone Science Laboratory, Hefei National Laboratory, Hefei 230088, China}
  
\author{Shiyan Li}
  \email{Contact author: shiyan$\_$li@fudan.edu.cn}
  \affiliation{State Key Laboratory of Surface Physics, Department of Physics, Fudan University, Shanghai 200438, China}
  \affiliation{Shanghai Research Center for Quantum Sciences, Shanghai 201315, China}
  \affiliation{Shanghai Branch, Hefei National Laboratory, Shanghai 201315, China}
  \affiliation{Collaborative Innovation Center of Advanced Microstructures, Nanjing 210093, China}

\date{\today}

\begin{abstract}

    We investigate the Seebeck coefficient ($S$) in infinite-layer nickelate films with different disorder levels. The disordered NdNiO$_{2}$ film exhibits a flat $S/T$ curve, whereas cleaner samples display a logarithmic divergence with decreasing temperature, followed by a pronounced ``hump'' near 25 K. These distinct behaviors reveal a disorder-driven transition from band-structure-dominated transport to quantum-critical-dominated transport. Below the ``hump'' temperature, four-fold symmetry breaking is observed in the in-plane angular magnetoresistance, indicating the presence of short-range antiferromagnetic order in parent infinite-layer nickelate films. Furthermore, the logarithmic divergence in $S/T$ is also observed in a clean superconducting Sm$_{0.73}$Ca$_{0.05}$Eu$_{0.22}$NiO$_{2}$ film, where it coexists with linear-in-temperature resistivity over the same temperature range. These findings demonstrate the existence of quantum criticality over a wide doping range in clean infinite-layer nickelate films, similar to cuprates, which highlights the central role of antiferromagnetic spin correlations in their superconducting pairing mechanisms. 

\end{abstract}

\maketitle

Upon doping Mott insulators, unconventional superconductivity was discovered in cuprates \cite{01Lee2006,02Armitage2010}. Building on this foundational principle, infinite-layer nickelates (Nd$_{1-x}$Sr$_{x}$NiO$_{2}$, or NSNO) were subsequently developed \cite{03Li2019}. The superconductivity of nickelates is realized through a topotactic transition from the perovskite phase NdNiO$_{3}$ via a CaH$_{2}$ reduction reaction combined with Sr doping. This new superconductor displays a crystal structure and superconducting dome similar to those observed in cuprates. Nevertheless, key distinctions exist, including the absence of long-range antiferromagnetic (AF) order in its parent compound NdNiO$_{2}$ \cite{04Lu2021,05Fowlie2022}, a Mott–Hubbard orbital alignment \cite{06Hepting2020,07Goodge2021}, and a multiband electronic structure \cite{08Botana2020,09Li2020}. These features introduce new challenges in elucidating the superconducting  mechanism of infinite-layer nickelates.

Recently, a linear-in-temperature resistivity has been reported on near optimal doping NSNO films as disorder decreases \cite{10Lee2023}. This linear resistivity, a hallmark of quantum critical behavior referred to as a ``strange metal'', has been observed in various unconventional superconductors, such as cuprates \cite{11Cooper2009,12Jin2011,13Yuan2021}, organic conductors \cite{14Doiron-Leyraud2009}, iron-based superconductors \cite{15Hayes2016,16Jiang2023}, and heavy-fermion superconductors \cite{17Park2008,18Hu2017,19Nguyen2021}. Thus, this finding in infinite-layer nickelates suggests a potential convergence in the quantum criticality among these unconventional superconducting families.

Thermoelectricity, as indicated by the Seebeck coefficient ($S$), is a traditional probe of electronic properties. The Seebeck coefficient is typically understood through Boltzmann transport theory. Under the elastic scattering approximation, the mean free path of all quasiparticles is energy-independent, therefore $S/T$ depends solely on the Fermi energy. Consequently, $S/T$ is used to study the band structure in a manner analogous to the Hall coefficient ($R$$_H$) in unconventional superconductors \cite{20Li2007,21Doiron-Leyraud2013}. Notably, in overdoped cuprates, $S/T$ deviates from band structure predictions \cite{22Jin2009,23Verret2017} and exhibits a positive value with logarithmic divergence (-$lnT$), which is associated with quantum criticality \cite{24Mandal2019,25Collignon2021}. Theoretical calculations suggest that this quantum critical behavior in the Seebeck coefficient emerges only when the effects of disorder are minimized \cite{26Grissonnanche2024}; otherwise, it becomes obscured by the features of band structure. If validated, these findings position the Seebeck coefficient as a valuable tool for investigating quantum criticality through disorder tuning in strongly correlated systems.

In this Letter, we investigate the $S/T$ in parent NdNiO$_{2}$ (NNO) and superconducting Sm$_{0.73}$Ca$_{0.05}$Eu$_{0.22}$NiO$_{2}$ (SCE$_{0.22}$) films. NNO films with different disorder levels exhibit markedly distinct Seebeck behaviors. The disordered NNO film maintains a temperature-independent negative $S/T$, indicating that electron carriers dominate charge transport. In contrast, the cleaner NNO films exhibit a logarithmic divergence in $S/T$, driven by energy-dependent inelastic scattering associated with quantum criticality. At lower temperatures, $S/T$ drops rapidly following the onset of four-fold symmetry breaking, suggesting the emergence of short-range AF order in clean NNO films. Furthermore, this logarithmic divergence in $S/T$ is also observed in the clean superconducting SCE$_{0.22}$ film, where it coexists with linear-in-temperature resistivity over the same temperature range. These observations indicate that quantum criticality persists over a wide doping range in clean infinite-layer nickelate films, similar to that observed in cuprates.

\begin{figure}
	\includegraphics[clip,width=8.5cm]{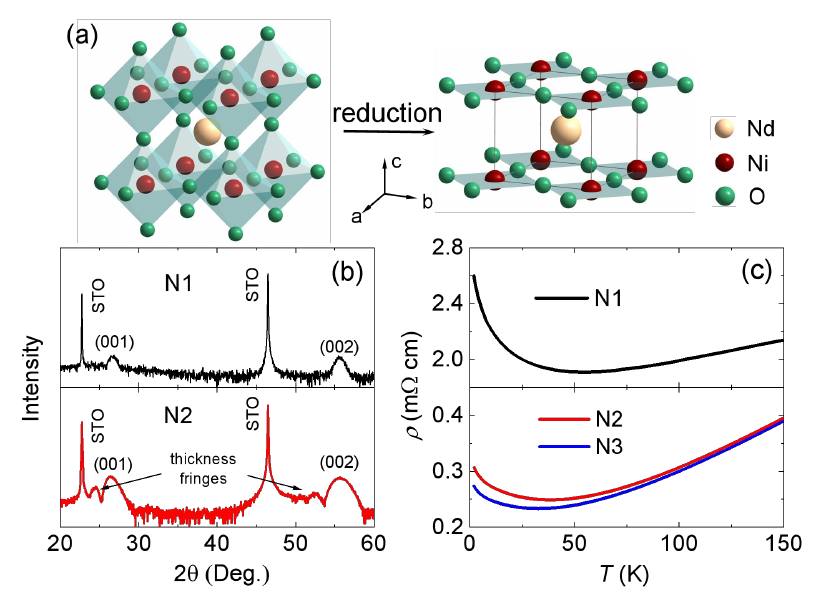}
	\caption{(a) Schematic illustration of the preparation of the infinite-layer NdNiO$_{2}$ films. (b) X-ray diffraction patterns of NdNiO$_{2}$ films, in which only (00$l$) Bragg peaks are observed. (c) Temperature-dependent resistivity of NdNiO$_{2}$ films.}
\end{figure}

NdNiO$_{3}$ and Sm$_{0.73}$Ca$_{0.05}$Eu$_{0.22}$NiO$_{3}$ thin films were deposited on SrTiO$_{3}$ (00$l$) and (LaAlO$_{3}$)$_{0.33}$(Sr$_{2}$TaAlO$_{6}$)$_{0.7}$ (00$l$) substrates using pulsed-laser deposition, respectively. Subsequently, infinite-layer thin films were obtained from the perovskite phase  via a topochemical reaction [Fig. 1(a)]. NNO sample 1 (N1) and SCE$_{0.22}$ sample (S1) were prepared using CaH$_{2}$ powders as reducing agents \cite{27Tam2022,28Yang2025}, while NNO sample 2 and 3 (N2 and N3) were produced utilizing a new method that employs atomic hydrogen as reducing agents \cite{29Ding2024,30Li2025}. The resistivity and Seebeck coefficient were measured using a Physical Property Measurement System (DynaCool, Quantum Design). For Seebeck experiment, a longitudinal thermal gradient of approximately $\Delta T \approx$ 1 K was established along the sample length, as measured by two Cernox-1050 thermometers. The corresponding Seebeck voltage $\Delta V$ was measured with phosphor-bronze wires using the same electrical contacts as those for $\Delta T$. The background voltage variation remained below 100 nV over the entire measurement temperature range for each sample, ensuring that the results were free of spurious signals. The Seebeck coefficient was calculated as -$\Delta V$/$\Delta T$.

Figure 1(b) presents the X-ray diffraction (XRD) patterns for N1 and N2, where only (00$l$) Bragg peaks are observed. N1 has a thickness of 10 nm, whereas N2 is 7.6 nm. The reduced thickness of N2 results in broader (00$l$) Bragg peaks. Clear thickness fringes in the XRD pattern of N2 indicate the successful fabrication of a high-quality infinite-layer phase \cite{10Lee2023,31Parzyck2024}. This is supported by the resistivity ($\rho$) data in Fig. 1(c), where N2 and N3 manifest weaker insulating behavior at low temperature than N1. The minimal resistivity of N2 is approximately 0.25 m$\Omega$ cm, an order of magnitude lower than the value for N1 (1.9 m$\Omega$ cm), and comparable to the best NNO films previously reported \cite{10Lee2023,32Parzyck2024}.

\begin{figure}
	\includegraphics[clip,width=7.0cm]{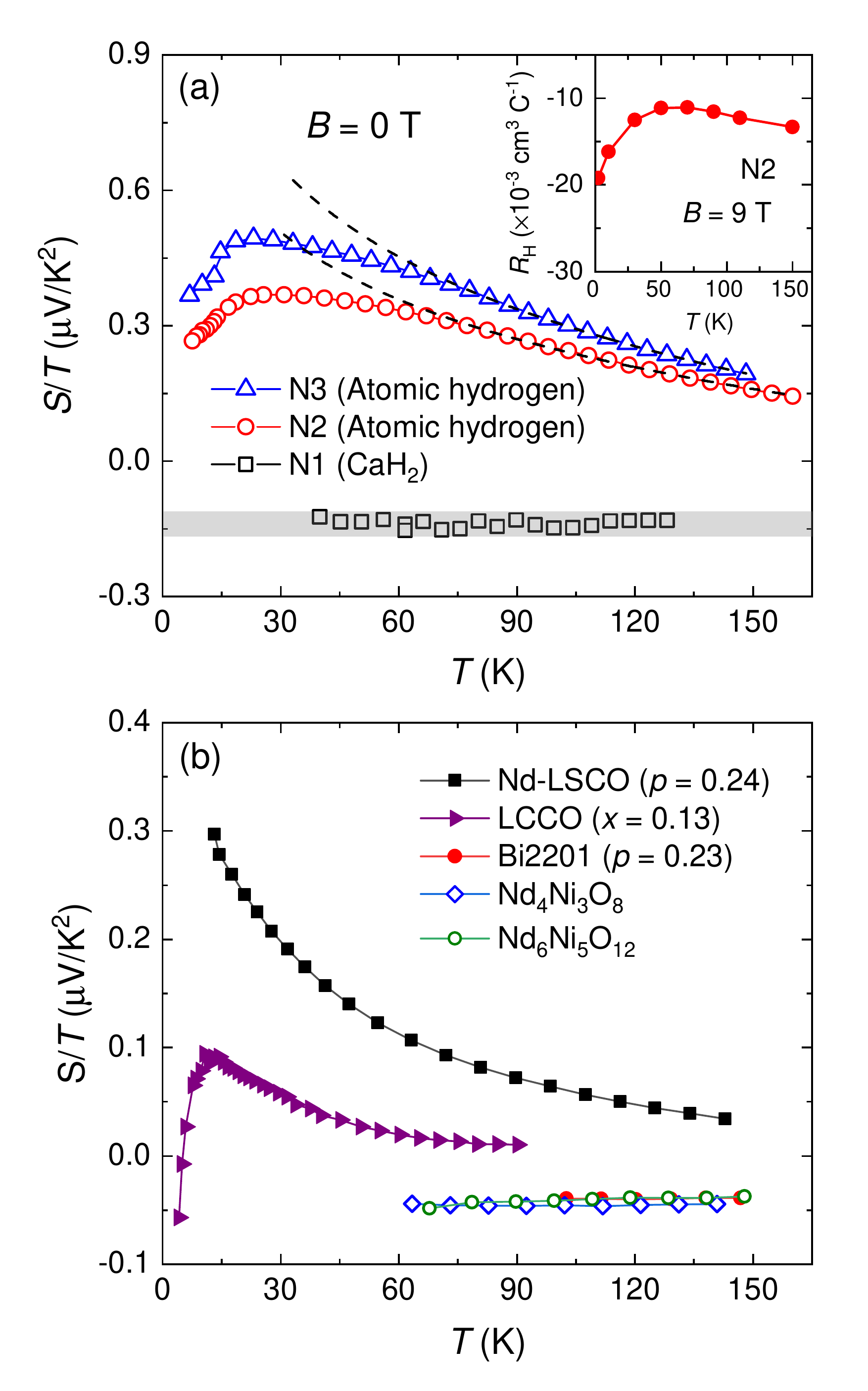}
	\caption{(a) The Seebeck coefficients of NdNiO$_{2}$ films at zero magnetic field, plotted as $S/T$ versus $T$. The dashed lines represent logarithmic fit using the function of $a-blnT$. A hump emerges at approximately 25 K for samples N2 and N3. Inset: The $R_H$ as a function of temperature for N2 at $B$ = 9 T. (b) The Seebeck coefficients in normal state for various materials include Nd-LSCO ($p = 0.24$, $B$ = 16 T) \cite{35Gourgout2022}, LCCO ($x = 0.13$, $B$ = 9 T) \cite{24Mandal2019}, Bi2201 ($p = 0.23$, $B$ = 0 T) \cite{36Kondo2005}, Nd$_{4}$Ni$_{3}$O$_{8}$ and Nd$_{6}$Ni$_{5}$O$_{12}$ ($B$ = 0 T) \cite{26Grissonnanche2024}.}
\end{figure}

Figure 2 shows the temperature dependence of $S/T$ for NNO thin films. Samples with varying levels of disorder exhibit markedly different Seebeck behaviors, with distinct signs and temperature dependencies. N1 exhibits a negative, nearly temperature-independent $S/T$, while N2 and N3 exhibit positive values. As the temperature decreases, the $S/T$ curves of N2 and N3 first diverge logarithmically, followed by a sudden drop that develops a ``hump'' feature near $T_{smax}$ = 25 K. This behavior resembles that observed in the normal state ($B$ = 9 T) of electron-doped cuprate La$_{2-x}$Ce$_{x}$CuO$_{4}$ (LCCO, $x = 0.13$), where the ``hump'' is attributed to short-range AF ordering, and the logarithmic divergence is associated with proximity of the AF quantum critical point \cite{24Mandal2019}.

The Seebeck coefficient can be expressed using the Mott formula under the elastic scattering approximation:
\begin{equation}
	S = - \frac{\pi^{2}}{3}\frac{Tk_{B}^{2}}{e}\frac{1}{\epsilon_{F}},
\end{equation}
where $e$ is the electron charge, $k_B$ is the Boltzmann constant, and $\epsilon$$_F$ is Fermi energy. This relation reflects the band structure, analogous to $R_H$ in unconventional superconductors \cite{20Li2007,21Doiron-Leyraud2013}. The temperature-independent negative $S/T$ observed in N1 is consistent with $R_H$ in the same temperature range (40 - 130 K) reported in previous studies \cite{03Li2019,09Li2020}, indicating that electron carriers dominate the transport properties. 

The significant changes in $S/T$ for N2 and N3 may arise from two possible origins: modifications of the band structure or breakdown of the Boltzmann equation due to inelastic scattering. According to XRD data, N1 and N2 have nearly identical crystal structures, with lattice constant $c$ differing by less than 0.3\% (3.305 {\AA} for N1, and 3.298 {\AA} for N2), therefore substantial differences in the band structure due to structural difference between them is unlikely. However, different reduction methods may result in different doping, which can shift the Fermi level and thereby alter the Fermi surfaces in nickelates. To further clarify the band structure, we measured the Hall effect of N2. As shown in the inset of Fig. 2(a), its $R_H$ remains negative and exhibits a temperature dependence similar to prior reports \cite{03Li2019,09Li2020}, ruling out band structure changes as the dominant factor.  

In this context, the significant changes in $S/T$ for N2 and N3 should come from the breakdown of the Boltzmann transport theory due to inelastic scattering. Similar breakdown has been reported in cuprates such as LCCO and La$_{1.6-x}$Nd$_{0.4}$Sr$_{x}$CuO$_{4}$ (Nd-LSCO) \cite{24Mandal2019,25Collignon2021}. Quantum critical behavior in $S/T$ $\propto$ $-lnT$ emerges over a wide doping range, which cannot be explained by band structure calculations \cite{24Mandal2019,25Collignon2021}. In cuprates, particularly in doping regimes governed by quantum criticality, the energy-dependent inelastic scattering rate has long been recognized, and is often described by the marginal Fermi liquid \cite{33Varma1989,34Varma2020}. Introducing an inelastic, energy-dependent scattering rate $1⁄{\tau(\epsilon)}$ reconciles theoretical predictions with experimental observations for Nd-LSCO ($p = 0.24$) \cite{35Gourgout2022}. In this framework, mass renormalization, expressed as $1⁄{Z(T)}$ $\propto$ $[1+log(\Lambda⁄T)]$, where $\Lambda$ is cutoff, amplifies the band (Mott) component to $S/T$, resulting in logarithmic divergence \cite{35Gourgout2022}.

Within above energy-dependent scattering scenario, the total scattering rate ($1⁄{\tau_{t}}$) is given by the sum of elastic and inelastic components:
\begin{equation}
	\frac{1}{\tau_{t}} = \frac{1}{\tau_0} + \frac{1}{\tau(\epsilon)},
\end{equation}
 where $1⁄\tau_0$ is the elastic scattering rate originating from disorder, and $1⁄\tau(\epsilon)$ is the inelastic scattering rate associated with quantum criticality \cite{35Gourgout2022}. The primary distinction among our NNO samples lies in the different levels of disorder. The appearance of thickness fringes in XRD, lower resistivity, and weaker upturn in the $\rho(T)$ curve confirm that N2 and N3 are significantly cleaner than N1. Further evidence comes from the observation of quantum-well states in NNO samples grown alongside N2, as revealed by angle-resolved photoemission spectroscopy (ARPES) measurements \cite{30Li2025}. Thus, N2 and N3 are much closer to the ``clean'' limit in comparison to N1. In N1, more disorders make $1⁄\tau_0 \gg 1⁄\tau(\epsilon)$, resulting in a predominantly energy-independent total scattering rate. This allows the Seebeck coefficient to be described by Eq. 1 and primarily reflects the band structure. As disorder is reduced, the inelastic scattering rate becomes dominant, driving the system to exhibit quantum critical behavior, as evidenced by the -$lnT$ dependence of $S/T$ for N2 and N3. 

Our findings suggest that the Seebeck coefficient is governed by the relative magnitudes of the elastic and inelastic scattering rates. Figure 2(b) presents the temperature dependence of $S/T$ for several cuprates and nickelates, which exhibit thermoelectric responses comparable to thoese of our NNO films. Nd-LSCO ($p = 0.24$) displays a positive logarithmic divergence in $S/T$, while (Bi,Pb)$_{2}$(Sr,La)$_{2}$CuO$_{6+\delta}$ (Bi2201, $p = 0.23$) exhibits temperature-independent behavior \cite{36Kondo2005}. Notably, the band structures of these two materials are similar. For analogous systems, disorder levels can be estimated by comparing resistivities. The residual resistivity $\rho_{0}$ of Bi2201 ($p = 0.23$) is approximately 120 $\mu$$\Omega$ cm \cite{37Ayres2021}, significantly larger than the $\rho_{0}$ of 23 $\mu$$\Omega$ cm for Nd-LSCO ($p = 0.24$) \cite{38Daou2009}. Consequently, Bi2201 ($p = 0.23$) has more disorders than Nd-LSCO ($p = 0.24$). G. Grissonnanche \textit{et al}. recalculated $S/T$ for Nd-LSCO ($p = 0.24$) by varying the elastic scattering from 10 to 3500 $ps^{-1}$ \cite{26Grissonnanche2024}. Their results show that $S/T$ transitions from a positive $-lnT$ dependence to a temperature-independent negative value as disorder increases. In nickelates with $n = 3$ and $n = 5$, reported resistivities at 50 K are 1.2 and 2.1 m$\Omega$ cm, respectively \cite{39Pan2022}\textemdash both an order of magnitude higher than those of our ``clean'' samples\textemdash placing them in the ``dirty'' limit like N1, consistent with their temperature-independent $S/T$ behavior. For different systems, resistivity is an inadequate physical quantity for comparing the magnitude of scattering rates, as it is also influenced by effective mass and carrier concentration. Moreover, these materials possess different correlation strengths, leading to variations in $1⁄\tau(\epsilon)$. Consequently, although the resistivities of ``clean'' infinite-layer nickelates and Bi2201 ($p = 0.23$) are comparable, their $S/T$ exhibits distinct characteristics: one demonstrates quantum criticality, while the other is determined solely by underlying band structure. 

\begin{figure}
  \includegraphics[clip,width=7.0cm]{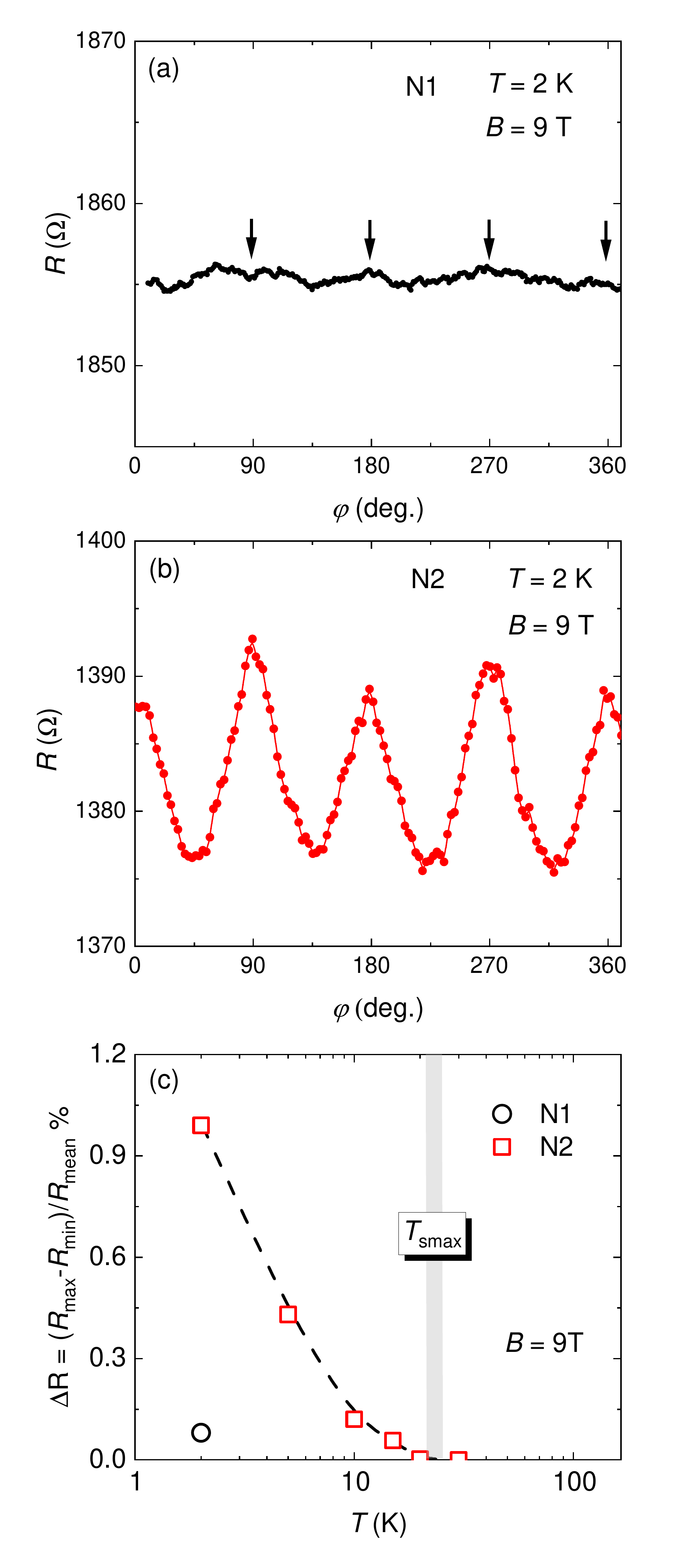}
  \caption{(a) and (b) The $R(\varphi)$ curves at 2 K and 9 T in the rectangular plot for N1 and N2. (c) The magnitude of the anisotropic in-plane AMR, defined as $\Delta$$R = [R_{max} - R_{min}] / R_{min}$, decreases with increasing temperature. $\Delta$$R$(9T) approaches zero around $T_{smax}$, indicating the disappearance of the four-fold AMR. The shadow represents the characteristic temperature of the hump behavior in the $S/T$ curve of N2.}
  \end{figure}

Given that thermoelectricity in ``clean'' NNO samples is governed by inelastic scattering rates, the emergence of ``hump'' behavior signifies a change in the scattering mechanism at low temperatures. Previous studies have reported Kondo scattering \cite{40Shao2023}, charge order \cite{27Tam2022}, and short-range AF order \cite{04Lu2021} in parent infinite-layer nickelate films, each contributing additional scattering. To elucidate the origin of the ``hump'' behavior, we performed in-plane angular magnetoresistance (AMR) measurements, which can probe magnetic order by examining spin-charge coupling \cite{22Jin2009,41Wu2008,42wang2009}. The NNO films were rotated around the $c$ axis with $B$ $||$ $ab$. N1 exhibits a very weak angle-dependent magnetoresistance [Fig. 3(a)], primarily due to the high disorder level present in N1, leading to a strong isotropic scattering rate. As shown in Fig. 3(b), a clearer four-fold symmetry emerges at 2 K under 9 T for N2. We define the magnitude of the anisotropic in-plane AMR as $\Delta$$R = [R_{max} - R_{min}] / R_{min}$, where $R_{max}$ and $R_{min}$ are the maximum and minimum resistance, respectively, as the rotation angle ($\varphi$) varies from 0$^{\circ}$ to 360$^{\circ}$. The anisotropic AMR magnitude is approximately 1\% at 2 K and decreases with increasing temperature, finally approaching zero around $T_{smax}$ [Fig. 3(c)]. This temperature-dependent $\Delta$$R$ indicates that the four-fold symmetry breaking dose not origin from lattice symmetry but rather from some kind of ordering, which creates a ``hump'' in $S/T$ curves at low temperatures.

Resonant soft X-ray scattering experiment has revealed a commensurate $3a_{0}$ charge-density-wave order in infinite-layer nickelate films \cite{27Tam2022}. However, subsequent measurements on a series of NdNiO$_{2+x}$ samples demonstrated that the absence of charge order in fully reduced single-phase NNO films \cite{31Parzyck2024}. Instead, the observed $3a_{0}$ superlattice peak arises from a partially reduced impurity phase where excess apical oxygens form ordered rows with three-unit-cell periodicity \cite{31Parzyck2024}. Consistently, our X-ray scattering results for samples grown alongside N2 confirmed the absence of charge order in high-quality films \cite{30Li2025}. Therefore, alternative explanations based on AF ordering merit consideration. Although the parent compound of infinite-layer nickelates lacks long-rang AF order \cite{05Fowlie2022,43Wu2024}, dispersive excitations with a bandwidth of approximately 200 meV have been observed in NNO films \cite{04Lu2021}, resembling spin waves in antiferromagnetically aligned spins on a square lattice. In electron-doped cuprates, long-range AF order disappears in the underdoped region \cite{44Motoyama2007,45Saadaoui2015}, while short-range AF order persists into the overdoped region \cite{02Armitage2010,46Mang2004,47Matsui2005,48Sarkar2017}. Within this regime, “hump” behavior and -$lnT$ divergence appear in $S/T$ curves, as shown in Fig. 2(b), accompanied by two-fold (for LCCO) and four-fold (for Nd$_{2-x}$Ce$_{x}$CuO$_{4}$) symmetry breaking in the in-plane AMR \cite{41Wu2008,22Jin2009}. These similar phenomena suggest the presence of short-range AF order in parent infinite-layer nickelate films.

According to ARPES results, the Fermi surfaces of NNO films consist of a quasi-two-dimensional (quasi-2D) hole-like $\alpha$ band and a 3D electron-like $\beta$ band \cite{30Li2025}. The negative Hall coefficient suggests that the electron band dominates in electrical transport. Corresponding negative values are also observed in Seebeck measurements on disordered samples, because the high elastic scattering limit of the Seebeck coefficient reflects only the underlying band structure, analogous to the Hall coefficient. In clean samples, inelastic scattering associated with AF spin correlations dominates. These AF spin correlations originate from the Ni $d_{x^{2}-y^{2}}$ orbital, which forms the hole-like $\alpha$ band \cite{30Li2025}. Consequently, a positive divergence of $S/T$ is observed in infinite-layer NNO films.
 
\begin{figure}
	\includegraphics[clip,width=8.5cm]{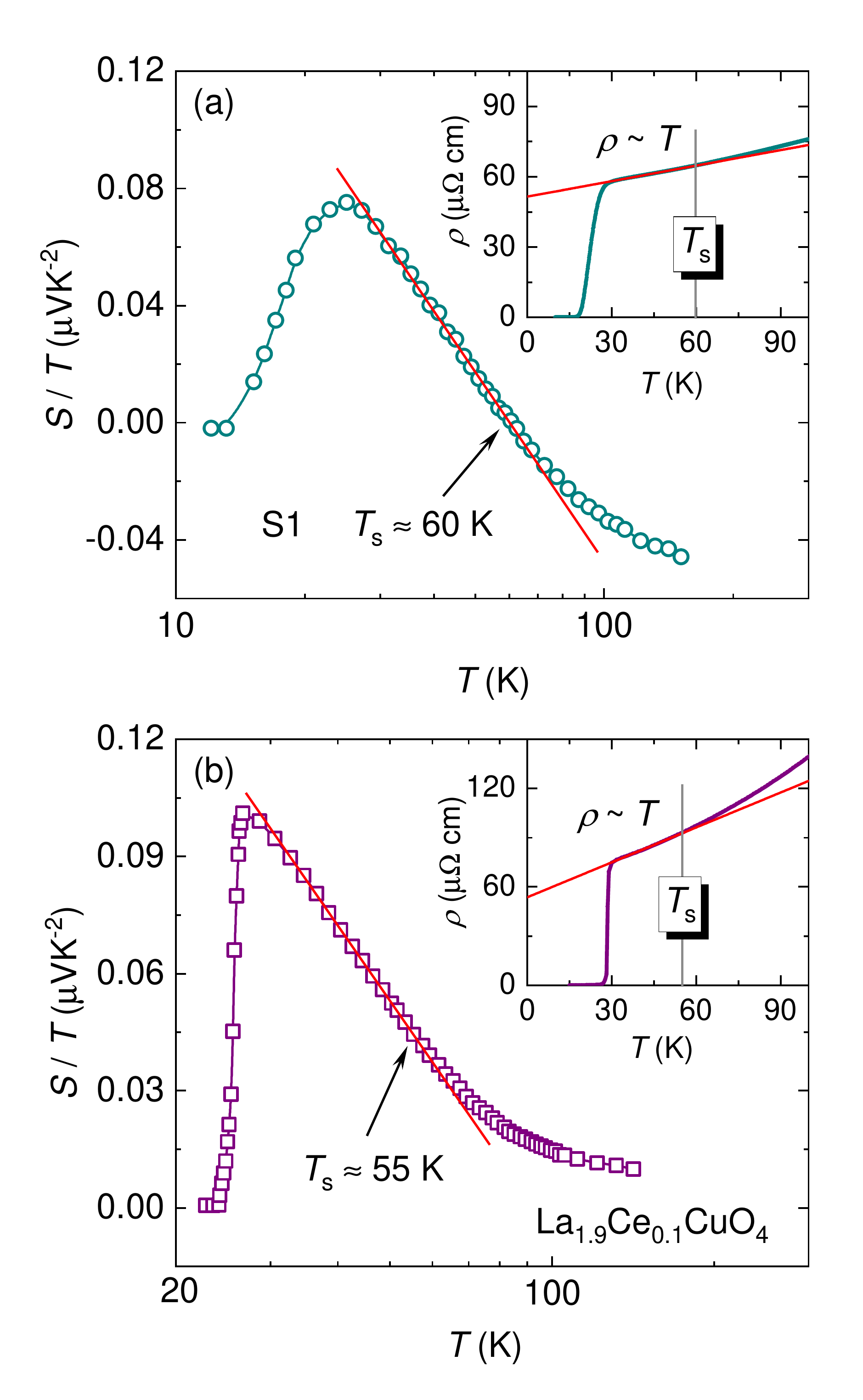}
	\caption{The $S/T$ of superconducting (a) SCE$_{0.22}$ and (b) LCCO \cite{49Zhang} films is plotted on a logarithmic scale. For both films, $S/T$ exhibits $–lnT$ dependence above $T_c$, up to the characteristic temperature $T_s$. In this temperature range, the linear resistivity is also observed. The red solid line is a guide to the eye.}
\end{figure}

Below, we demonstrate that quantum criticality also emerges in the clean superconducting SCE$_{0.22}$ film, which exhibits a low residual resistivity of only 51 $\mu$$\Omega$ cm, suggesting minimal disorder scattering. As shown in Fig. 4(a), $S/T$ follows a $–lnT$ dependence above the superconducting transition temperature ($T_c$) and deviates at $T_s$. Furthermore, the $S/T$ of the SCE$_{0.22}$ film transitions from negative to positive as the temperature decreases, while its $R_H$ remains consistently negative \cite{28Yang2025}. This discrepancy is a typical characteristic of quantum-critical-dominated transport, as discussed above for clean NNO films. The linear resistivity, known as the ``strange'' metal state, is commonly associated with quantum criticality \cite{10Lee2023}. Within this temperature range between $T_c$ and $T_s$, the resistivity of SCE$_{0.22}$ film remains strictly linear in $T$ [inset of Fig. 4(a)], paralleling that of the optimally doped LCCO ($x = 0.1$) shown in Fig. 4(b) \cite{49Zhang}. This coexistence of linear resistivity and logarithmic divergence on $S/T$ not only persists across a wide doping range in both electron-doped cuprate LCCO and hole-doped cuprate Nd-LSCO \cite{25Collignon2021,38Daou2009,50Legros}, but has also been reported in other unconventional superconductors, including iron-based superconductor Fe$_{y}$Se$_{1-x}$Te$_{x}$ \cite{51Juan}, heavy-fermion superconductor YbRh$_{2}$Si$_{2}$ \cite{52Hartmann,53Machida,19Nguyen2021}, and organic superconductor $\kappa$-(BEDT-TTF)$_{4}$Hg$_{2.89}$Br$_{8}$ \cite{54Okuhata,55Wakamatsu}. These findings indicate a convergence of quantum critical behavior among diverse families of unconventional superconductors.

The origin of superconductivity in infinite-layer nickelates remains an open question. Our results reveal the present of short-range AF order and quantum critical behavior similar to other unconventional superconductors, especially to cuprates, thereby strongly supporting the AF-fluctuation–mediated pairing scenario in infinite-layer nickelates. Previous studies have also reported four-fold symmetry breaking in the mixed states of optimally doped NSNO \cite{56Ji2023}, suggesting that AF spin superexchange coupling persists at high doping and favors $d$-wave superconductivity in infinite-layer nickelate films, supported by scanning tunneling spectroscopy and ultrafast terahertz spectroscopy results \cite{57Gu2020,58Cheng2024}. 

In summary, we investigated the Seebeck coefficient of NNO and SCE$_{0.22}$ films. By comparing NNO films with different disorder levels, we find that the inelastic scattering rate associated with quantum criticality becomes prominent as disorder decreases, leading to an $S/T \propto -lnT$ relationship in the ``clean'' films. This logarithmic divergence is also present in the clean superconducting SCE$_{0.22}$ film, where it coexists with the linear resistivity within the same temperature range. These findings emphasize the important role of AF spin fluctuations in infinite-layer nickelates and suggest a convergence of quantum critical behavior among unconventional superconducting families as disorder decreases. Finally, our results demonstrate that disorder-tunable thermoelectric measurement provides an effective approach for probing quantum criticality and band structure in strongly correlated systems.
 
This work was supported by the Natural Science Foundation of China (Grants No. 12174064, No. 12034004, No. 92477206, No. 12174325, No. 12274061, and No. 12304078), the Shanghai Municipal Science and Technology Major Project (Grant No. 2019SHZDZX01), the Innovation Program for Quantum Science and Technology (Grants No. 2024ZD0300104 and No. 2021ZD0302803), the New Cornerstone Science Foundation, the Key Research and Development Program from the Ministry of Science and Technology (2023YFA1406301), the Science and Technology Department of Sichuan Province (Grant No. 2024ZYD0164), and a Guangdong Basic and Applied Basic Research Grant (Grant No. 2023A1515011352). This work was also supported by the Research Grants Council (RGC) of the Hong Kong Special Administrative Region, China, under Early Career Scheme and General Research Fund (CityU 21301221, CityU 11309622, CityU 11300923, CityU 11313325).


\begin{thebibliography}{99}
\bibitem{01Lee2006}	P. A. Lee, N. Nagaosa, and X.-G. Wen, Doping a Mott insulator: Physics of high-temperature superconductivity,
\href{https://journals.aps.org/rmp/abstract/10.1103/RevModPhys.78.17}{Rev. Mod. Phys. {\bf78}, 17 (2006)}.
\bibitem{02Armitage2010} N. P. Armitage, P. Fournier, and R. L. Greene, Progress and perspectives on electron-doped cuprates,
\href{https://journals.aps.org/rmp/abstract/10.1103/RevModPhys.82.2421}{Rev. Mod. Phys. {\bf82}, 2421 (2010)}.
\bibitem{03Li2019}	D. Li, K. Lee, B. Y. Wang, M. Osada, S. Crossley, H. R. Lee, Y. Cui, Y. Hikita, and H. Y. Hwang, Superconductivity in an infinite-layer nickelate,
\href{https://www.nature.com/articles/s41586-019-1496-5}{Nature {\bf572}, 624 (2019)}.
\bibitem{04Lu2021}	H. Lu, M. Rossi, A. Nag, M. Osada, D.F. Li, K. Lee, B. Y. Wang, M. Garcia-Fernandez, S. Agrestini, Z. X. Shen, E. M. Been, B. Moritz, T. P. Devereaux, J. Zaanen, H. Y. Hwang, K.-J. Zhou, and W. S. Lee, Magnetic excitations in infinite-layer nickelates,
\href{https://www.science.org/doi/10.1126/science.abd7726}{Science {\bf373}, 213 (2021)}.
\bibitem{05Fowlie2022} J. Fowlie, M. Hadjimichael, M. M. Martins, D. Li, M. Osada, B. Y. Wang, K. Lee, Y. Lee, Z. Salman, T. Prokscha, J.-M. Triscone, H. Y. Hwang, and A. Suter, Intrinsic magnetism in superconducting infinite-layer nickelates,
\href{https://www.nature.com/articles/s41567-022-01684-y}{Nat. Phys. {\bf18}, 1043 (2022)}.
\bibitem{06Hepting2020} M. Hepting, D. Li, C. J. Jia, H. Lu, E. Paris, Y. Tseng, X. Feng, M. Osada, E. Been, Y. Hikita, Y. D. Chuang, Z. Hussain, K. J. Zhou, A. Nag, M. Garcia-Fernandez, M. Rossi, H. Y. Huang, D. J. Huang, Z. X. Shen, T. Schmitt, H. Y. Hwang, B. Moritz, J. Zaanen, T. P. Devereaux, and W. S. Lee, Electronic structure of the parent compound of superconducting infinite-layer nickelates,
\href{https://www.nature.com/articles/s41563-019-0585-z}{Nat. Mater. {\bf19}, 381 (2020)}.
\bibitem{07Goodge2021} B. H. Goodge, D. Li, K. Lee, M. Osada, B. Y. Wang, G. A. Sawatzky, H. Y. Hwang, and L. F. Kourkoutis, Doping evolution of the Mott–Hubbard landscape in infinite-layer nickelates,
\href{https://www.pnas.org/doi/10.1073/pnas.2007683118}{Proc. Natl. Acad. Sci. U. S. A. {\bf118}, e2007683118 (2021)}.
\bibitem{08Botana2020} A. S. Botana and M. R. Norman, Similarities and differences between LaNiO$_{2}$ and CaCuO$_{2}$ and implications for superconductivity,
\href{https://journals.aps.org/prx/abstract/10.1103/PhysRevX.10.011024}{Phys. Rev. X {\bf10}, 011024 (2020)}.
\bibitem{09Li2020} D. Li, B. Y. Wang, K. Lee, S. P. Harvey, M. Osada, B. H. Goodge, L. F. Kourkoutis, and H. Y. Hwang, Superconducting dome in Nd$_{1-x}$Sr$_{x}$NiO$_{2}$ infinite layer films,
\href{https://journals.aps.org/prl/abstract/10.1103/PhysRevLett.125.027001}{Phys. Rev. Lett. {\bf125}, 027001 (2020)}.
\bibitem{10Lee2023} K. Lee, B. Y. Wang, M. Osada, B. H. Goodge, T. C. Wang, Y. Lee, S. Harvey, W. J. Kim, Y. Yu, C. Murthy, S. Raghu, L. F. Kourkoutis, and H. Y. Hwang, Linear-in-temperature resistivity for optimally superconducting (Nd,Sr)NiO$_{2}$,
\href{https://www.nature.com/articles/s41586-023-06129-x}{Nature {\bf619}, 288 (2023)}.
\bibitem{11Cooper2009} R. A. Cooper, Y. Wang, B. Vignolle, O. J. Lipscombe, S. M. Hayden, Y. Tanabe, T. Adachi, Y. Koike, M. Nohara, H. Takagi, C. Proust, and N. E. Hussey, Anomalous criticality in the electrical resistivity of La$_{2-x}$Sr$_{x}$CuO$_{4}$,
\href{https://www.science.org/doi/10.1126/science.1165015}{Science {\bf323}, 603 (2009)}.
\bibitem{12Jin2011} K. Jin, N. P. Butch, K. Kirshenbaum, J. Paglione, and R. L. Greene, Link between spin fluctuations and electron pairing in copper oxide superconductors,
\href{https://www.nature.com/articles/nature10308}{Nature {\bf476}, 73 (2011)}.
\bibitem{13Yuan2021} J. Yuan, Q. Chen, K. Jiang, Z. Feng, Z. Lin, H. Yu, G. He, J. Zhang, X. Jiang, X. Zhang, Y. Shi, Y. Zhang, M. Qin, Z. G. Cheng, N. Tamura, Y.-f. Yang, T. Xiang, J. Hu, I. Takeuchi, K. Jin, and Z. Zhao, Scaling of the strange-metal scattering in unconventional superconductors,
\href{https://www.nature.com/articles/s41586-021-04305-5}{Nature {\bf602}, 431 (2022)}.
\bibitem{14Doiron-Leyraud2009} N. Doiron-Leyraud, P. Auban-Senzier, S. René de Cotret, C. Bourbonnais, D. Jérome, K. Bechgaard, and L. Taillefer, Correlation between linear resistivity and $T$$_{c}$ in the Bechgaard salts and the pnictide superconductor Ba(Fe$_{1-x}$Co$_{x}$)$_{2}$As$_{2}$,
\href{https://journals.aps.org/prb/abstract/10.1103/PhysRevB.80.214531}{Phys. Rev. B {\bf80}, 214531 (2009)}.
\bibitem{15Hayes2016} I. M. Hayes, R. D. McDonald, N. P. Breznay, T. Helm, P. J. W. Moll, M. Wartenbe, A. Shekhter, and J. G. Analytis, Scaling between magnetic field and temperature in the high-temperature superconductor BaFe$_{2}$(As$_{1-x}$P$_{x}$)$_{2}$,
\href{https://www.nature.com/articles/nphys3773}{Nat. Phys. {\bf12}, 916 (2016)}.
\bibitem{16Jiang2023} X. Jiang, M. Qin, X. Wei, L. Xu, J. Ke, H. Zhu, R. Zhang, Z. Zhao, Q. Liang, Z. Wei, Z. Lin, Z. Feng, F. Chen, P. Xiong, J. Yuan, B. Zhu, Y. Li, C. Xi, Z. Wang, M. Yang, J. Wang, T. Xiang, J. Hu, K. Jiang, Q. Chen, K. Jin, and Z. Zhao, Interplay between superconductivity and the strange-metal state in FeSe,
\href{https://www.nature.com/articles/s41567-022-01894-4}{Nat. Phys. {\bf19}, 365 (2023)}.
\bibitem{17Park2008} T. Park, V. A. Sidorov, F. Ronning, J.-X. Zhu, Y. Tokiwa, H. Lee, E. D. Bauer, R. Movshovich, J. L. Sarrao, and J. D. Thompson, Isotropic quantum scattering and unconventional superconductivity,
\href{https://www.nature.com/articles/nature07431}{Nature {\bf456}, 366 (2008)}.
\bibitem{18Hu2017} T. Hu, Y. Liu, H. Xiao, G. Mu, and Y.-f. Yang, Universal linear-temperature resistivity: possible quantum diffusion transport in strongly correlated superconductors,
\href{https://www.nature.com/articles/s41598-017-09792-z}{Sci. Rep. {\bf7}, 9469 (2017)}.
\bibitem{19Nguyen2021} D. H. Nguyen, A. Sidorenko, M. Taupin, G. Knebel, G. Lapertot, E. Schuberth, and S. Paschen, Superconductivity in an extreme strange metal,
\href{https://www.nature.com/articles/s41467-021-24670-z}{Nat. Commun. {\bf12}, 4341 (2021)}.
\bibitem{20Li2007} P. Li, K. Behnia, and R. L. Greene, Evidence for a quantum phase transition in electron-doped Pr$_{2-x}$Ce$_{x}$CuO$_{4-\delta}$ from thermopower measurements, 
\href{https://journals.aps.org/prb/abstract/10.1103/PhysRevB.75.020506}{Phys. Rev. B {\bf75}, 020506(R) (2007)}.
\bibitem{21Doiron-Leyraud2013} N. Doiron-Leyraud, S. Lepault, O. Cyr-Choinière, B. Vignolle, G. Grissonnanche, F. Laliberté, J. Chang, N. Barišić, M. K. Chan, L. Ji, X. Zhao, Y. Li, M. Greven, C. Proust, and L. Taillefer, Hall, Seebeck, and Nernst coefficients of underdoped HgBa$_{2}$CuO$_{4+\delta}$: Fermi-surface reconstruction in an archetypal cuprate superconductor,
\href{https://journals.aps.org/prx/abstract/10.1103/PhysRevX.3.021019}{Phys. Rev. X {\bf3}, 021019 (2013)}.
\bibitem{22Jin2009} K. Jin, X. H. Zhang, P. Bach, and R. L. Greene, Evidence for antiferromagnetic order in La$_{2-x}$Ce$_{x}$CuO$_{4}$ from angular magnetoresistance measurements, 
\href{https://journals.aps.org/prb/abstract/10.1103/PhysRevB.80.012501}{Phys. Rev. B {\bf80}, 012501 (2009)}.
\bibitem{23Verret2017} S. Verret, O. Simard, M. Charlebois, D. Sénéchal, and A.-M. S. Tremblay, Phenomenological theories of the low-temperature pseudogap: Hall number, specific heat, and Seebeck coefficient,
\href{https://journals.aps.org/prb/abstract/10.1103/PhysRevB.96.125139}{Phys. Rev. B {\bf96}, 125139 (2017)}.
\bibitem{24Mandal2019} P. R. Mandal, T. Sarkar, and R. L. Greene, Anomalous quantum criticality in the electron-doped cuprates,
\href{https://www.pnas.org/doi/full/10.1073/pnas.1817653116}{Proc. Natl. Acad. Sci. U. S. A. {\bf116}, 5991 (2019)}.
\bibitem{25Collignon2021} C. Collignon, A. Ataei, A. Gourgout, S. Badoux, M. Lizaire, A. Legros, S. Licciardello, S. Wiedmann, J. Q. Yan, J. S. Zhou, Q. Ma, B. D. Gaulin, N. Doiron-Leyraud, and L. Taillefer, Thermopower across the phase diagram of the cuprate La$_{1.6-x}$Nd$_{0.4}$Sr$_{x}$CuO$_{4}$: Signatures of the pseudogap and charge density wave phases,
\href{https://journals.aps.org/prb/abstract/10.1103/PhysRevB.103.155102}{Phys. Rev. B {\bf103}, 155102 (2021)}.
\bibitem{26Grissonnanche2024} G. Grissonnanche, G. A. Pan, H. LaBollita, D. F. Segedin, Q. Song, H. Paik, C. M. Brooks, E. Beauchesne-Blanchet, J. L. Santana González, A. S. Botana, J. A. Mundy, and B. J. Ramshaw, Electronic band structure of a superconducting nickelate probed by the Seebeck coefficient in the disordered limit,
\href{https://journals.aps.org/prx/abstract/10.1103/PhysRevX.14.041021}{Phys. Rev. X {\bf14}, 041021 (2024)}.
\bibitem{27Tam2022} C. C. Tam, J. Choi, X. Ding, S. Agrestini, A. Nag, M. Wu, B. Huang, H. Luo, P. Gao, M. García-Fernández, L. Qiao, and K.-J. Zhou, Charge density waves in infinite-layer NdNiO$_{2}$ nickelates,
\href{https://www.nature.com/articles/s41563-022-01330-1}{Nat. Mater. {\bf21}, 1116 (2022)}.
\bibitem{28Yang2025} M. Yang, H. Wang, J. Tang, J. Luo, X. Wu, R. Mao, W. Xu, G. Zhou, Z. Dong, B. Feng, L. Shi, Z. Pei, P. Gao, Z. Chen, and D. Li, Enhanced superconductivity in co-doped infinite-layer samarium nickelate thin films,
\href{https://arxiv.org/pdf/2503.18346}{arXiv: 2503.18346}.
\bibitem{29Ding2024} X. Ding, Y. Fan, X. Wang, C. Li, Z. An, J. Ye, S. Tang, M. Lei, X. Sun, N. Guo, Z. Chen, S. Sangphet, Y. Wang, H. Xu, R. Peng, and D. Feng, Cuprate-like electronic structures in infinite-layer nickelates with substantial hole dopings,
\href{https://www.sciengine.com/NSR/doi/10.1093/nsr/nwae194}{Nat. Sci. Rev. {\bf11}, nwae194 (2024)}.
\bibitem{30Li2025} C. Li, Y. Chen, X. Ding, Y. Zhuang, N. Guo, Z. Chen, Y. Fan, J. Ye, Z. An, S. Sangphet, S. Tang, X. Wang, H. Huang, H. Xu, R. Peng, and D. Feng, Observation of electride-like $s$ states coexisting with correlated $d$ electrons in NdNiO$_{2}$,
\href{https://arxiv.org/pdf/2507.04378}{arXiv: 2507.04378, accepted by Phys. Rev. Lett.}.
\bibitem{31Parzyck2024} C. T. Parzyck, V. Anil, Y. Wu, B. H. Goodge, M. Roddy, L. F. Kourkoutis, D. G. Schlom, and K. M. Shen, Synthesis of thin film infinite-layer nickelates by atomic hydrogen reduction: Clarifying the role of the capping layer,
\href{https://pubs.aip.org/aip/apm/article/12/3/031132/3279003/Synthesis-of-thin-film-infinite-layer-nickelates}{APL Mater. {\bf12}, 031132 (2024)}.
\bibitem{32Parzyck2024}	C. T. Parzyck, N. K. Gupta, Y. Wu, V. Anil, L. Bhatt, M. Bouliane, R. Gong, B. Z. Gregory, A. Luo, R. Sutarto, F. He, Y. D. Chuang, T. Zhou, G. Herranz, L. F. Kourkoutis, A. Singer, D. G. Schlom, D. G. Hawthorn, and K. M. Shen, Absence of 3$a$$_{0}$ charge density wave order in the infinite-layer nickelate NdNiO$_{2}$,
\href{https://www.nature.com/articles/s41563-024-01797-0}{Nat. Mater. {\bf23}, 486 (2024)}.
\bibitem{33Varma1989} C. M. Varma, P. B. Littlewood, S. Schmitt-Rink, E. Abrahams, and A. E. Ruckenstein, Phenomenology of the normal state of Cu-O high-temperature superconductors,
\href{https://journals.aps.org/prl/abstract/10.1103/PhysRevLett.63.1996}{Phys. Rev. Lett. {\bf63}, 1996 (1989)}.
\bibitem{34Varma2020} C. M. Varma, Colloquium: Linear in temperature resistivity and associated mysteries including high temperature superconductivity,
\href{https://journals.aps.org/rmp/abstract/10.1103/RevModPhys.92.031001}{Rev. Mod. Phys. {\bf92}, 031001 (2020)}.
\bibitem{35Gourgout2022} A. Gourgout, G. Grissonnanche, F. Laliberté, A. Ataei, L. Chen, S. Verret, J. S. Zhou, J. Mravlje, A. Georges, N. Doiron-Leyraud, and L. Taillefer, Seebeck coefficient in a cuprate superconductor: Particle-hole asymmetry in the strange metal phase and Fermi surface transformation in the pseudogap phase,
\href{https://journals.aps.org/prx/abstract/10.1103/PhysRevX.12.011037}{Phys. Rev. X {\bf12}, 011037 (2022)}.
\bibitem{36Kondo2005} T. Kondo, T. Takeuchi, U. Mizutani, T. Yokoya, S. Tsuda, and S. Shin, Contribution of electronic structure to thermoelectric power in (Bi,Pb)$_{2}$(Sr,La)$_{2}$CuO$_{6+\delta}$,
\href{https://journals.aps.org/prb/abstract/10.1103/PhysRevB.72.024533}{Phys. Rev. B {\bf72}, 024533 (2005)}.
\bibitem{37Ayres2021}	J. Ayres, M. Berben, M. Čulo, Y. T. Hsu, E. van Heumen, Y. Huang, J. Zaanen, T. Kondo, T. Takeuchi, J. R. Cooper, C. Putzke, S. Friedemann, A. Carrington, and N. E. Hussey, Incoherent transport across the strange-metal regime of overdoped cuprates,
\href{https://www.nature.com/articles/s41586-021-03622-z}{Nature {\bf595}, 661 (2021)}.
\bibitem{38Daou2009}	R. Daou, N. Doiron-Leyraud, D. LeBoeuf, S. Y. Li, F. Laliberté, O. Cyr-Choinière, Y. J. Jo, L. Balicas, J. Q. Yan, J. S. Zhou, J. B. Goodenough, and L. Taillefer, Linear temperature dependence of resistivity and change in the Fermi surface at the pseudogap critical point of a high-$T$$_c$ superconductor,
\href{https://www.nature.com/articles/nphys1109}{Nat. Phys. {\bf5}, 31 (2009)}.
\bibitem{39Pan2022}	G. A. Pan, D. Ferenc Segedin, H. LaBollita, Q. Song, E. M. Nica, B. H. Goodge, A. T. Pierce, S. Doyle, S. Novakov, D. Córdova Carrizales, A. T. N’Diaye, P. Shafer, H. Paik, J. T. Heron, J. A. Mason, A. Yacoby, L. F. Kourkoutis, O. Erten, C. M. Brooks, A. S. Botana, and J. A. Mundy, Superconductivity in a quintuple-layer square-planar nickelate,
\href{https://www.nature.com/articles/s41563-021-01142-9}{Nat. Mater. {\bf21}, 160 (2022)}.
\bibitem{40Shao2023}	T.-N. Shao, Z.-T. Zhang, Y.-J. Qiao, Q. Zhao, H.-W. Liu, X.-X. Chen, W.-M. Jiang, C.-L. Yao, X.-Y. Chen, M.-H. Chen, R.-F. Dou, C.-M. Xiong, G.-M. Zhang, Y.-F. Yang, and J.-C. Nie, Kondo scattering in underdoped Nd$_{1-x}$Sr$_{x}$NiO$_{2}$ infinite-layer superconducting thin films,
\href{https://academic.oup.com/nsr/article/10/11/nwad112/7143129}{Nat. Sci. Rev. {\bf10}, nwad112 (2023)}.
\bibitem{41Wu2008}	T. Wu, C. H. Wang, G. Wu, D. F. Fang, J. L. Luo, G. T. Liu, and X. H. Chen, Giant anisotropy of the magnetoresistance and the `spin valve' effect in antiferromagnetic Nd$_{2-x}$Ce$_{x}$CuO$_{4}$, 
\href{https://iopscience.iop.org/article/10.1088/0953-8984/20/27/275226}{J. Phys. Condens. Matter {\bf20}, 275226 (2008)}.
\bibitem{42wang2009} X. F. Wang, T. Wu, G. Wu, R. H. Liu, H. Chen, Y. L. Xie, and X. H. Chen, The peculiar physical properties and phase diagram of BaFe$_{2-x}$Co$_{x}$As$_{2}$ single crystals, 
\href{https://iopscience.iop.org/article/10.1088/1367-2630/11/4/045003}{New J. Phys. {\bf11}, 045003 (2009)}.
\bibitem{43Wu2024} Q. Wu, Y. Fu, L. Wang, X. Zhou, S. Wang, Z. Zhu, K. Chen, C. Jiang, T. Shiroka, A. D. Hillier, J.-W. Mei, and L. Shu, Microscopic magnetism of nickel-based infinite-layer superconducting parent compounds $R$NiO$_{2}$ ($R$ = La, Nd): A $\mu$SR study,
\href{https://cpl.iphy.ac.cn/article/doi/10.1088/0256-307X/41/9/097502}{Chin. Phys. Lett. {\bf41}, 097502 (2024)}.
\bibitem{44Motoyama2007} E. M. Motoyama, G. Yu, I. M. Vishik, O. P. Vajk, P. K. Mang, and M. Greven, Spin correlations in the electron-doped high-transition-temperature superconductor Nd$_{2-x}$Ce$_{x}$CuO$_{4\pm\delta}$, 
\href{https://www.nature.com/articles/nature05437}{Nature {\bf445}, 186 (2007)}.
\bibitem{45Saadaoui2015}	H. Saadaoui, Z. Salman, H. Luetkens, T. Prokscha, A. Suter, W. A. MacFarlane, Y. Jiang, K. Jin, R. L. Greene, E. Morenzoni, and R. F. Kiefl, The phase diagram of electron-doped La$_{2-x}$Ce$_{x}$CuO$_{4-\delta}$,
\href{https://www.nature.com/articles/ncomms7041}{Nat. Commun. {\bf6}, 6041 (2015)}.
\bibitem{46Mang2004} P. K. Mang, O. P. Vajk, A. Arvanitaki, J. W. Lynn, and M. Greven, Spin correlations and magnetic order in nonsuperconducting Nd$_{2-x}$Ce$_{x}$CuO$_{4\pm\delta}$,
\href{https://journals.aps.org/prl/abstract/10.1103/PhysRevLett.93.027002}{Phys. Rev. Lett. {\bf93}, 027002 (2004)}.
\bibitem{47Matsui2005} H. Matsui, K. Terashima, T. Sato, T. Takahashi, S. C. Wang, H. B. Yang, H. Ding, T. Uefuji, and K. Yamada, Angle-resolved photoemission spectroscopy of the antiferromagnetic superconductor Nd$_{1.87}$Ce$_{0.13}$CuO$_{4}$: Anisotropic spin-correlation gap, pseudogap, and the induced quasiparticle mass enhancement,
\href{https://journals.aps.org/prl/abstract/10.1103/PhysRevLett.94.047005}{Phys. Rev. Lett. {\bf94}, 047005 (2005)}.
\bibitem{48Sarkar2017} T. Sarkar, P. R. Mandal, J. S. Higgins, Y. Zhao, H. Yu, K. Jin, and R. L. Greene, Fermi surface reconstruction and anomalous low-temperature resistivity in electron-doped La$_{2-x}$Ce$_{x}$CuO$_{4}$,
\href{https://journals.aps.org/prb/abstract/10.1103/PhysRevB.96.155449}{Phys. Rev. B {\bf96}, 155449 (2017)}.
\bibitem{49Zhang}	X. Zhang, H. Yu, J. Yuan, and K. Jin, unpublished. 
\bibitem{50Legros} A. Legros, S. Benhabib, W. Tabis, F. Laliberté, M. Dion, M. Lizaire, B. Vignolle, D. Vignolles, H. Raffy, Z. Z. Li, P. Auban-Senzier, N. Doiron-Leyraud, P. Fournier, D. Colson, L. Taillefer, and C. Proust, Universal $T$-linear resistivity and Planckian dissipation in overdoped cuprates, 
\href{https://www.nature.com/articles/s41567-018-0334-2}{Nat. Phys. {\bf15}, 142 (2019)}.
\bibitem{51Juan} J. Xu, M. Qin, Z. Lin, X. Zhang, R. Zhang, L. Xu, L. Zhang, Q. Shi, J. Yuan, B. Zhu, C. Dong, R. Xiong, Q. Chen, Y. Li, J. Shi, and K. Jin, \textit{in situ} electrical and thermal transport properties of Fe$_{y}$Se$_{1-x}$Te$_{x}$ films with ionic liquid gating, 
\href{https://journals.aps.org/prb/abstract/10.1103/PhysRevB.107.094514}{Phys. Rev. B {\bf107}, 094514 (2023)}.
\bibitem{52Hartmann} S. Hartmann, N. Oeschler, C. Krellner, C. Geibel, S. Paschen, and F. Steglich, Thermopower evidence for an abrupt Fermi surface change at the quantum critical point of YbRh$_{2}$Si$_{2}$, 
\href{https://journals.aps.org/prl/abstract/10.1103/PhysRevLett.104.096401}{Phys. Rev. Lett. {\bf104}, 096401 (2010)}.
\bibitem{53Machida} Y. Machida, K. Tomokuni, C. Ogura, K. Izawa, K. Kuga, S. Nakatsuji, G. Lapertot, G. Knebel, J.-P. Brison, and J. Flouquet, Thermoelectric response near a quantum critical point of $\beta$-YbAlB$_{4}$ and YbRh$_{2}$Si$_{2}$: A comparative study, 
\href{https://journals.aps.org/prl/abstract/10.1103/PhysRevLett.109.156405}{Phys. Rev. Lett. {\bf109}, 156405 (2012)}.
\bibitem{54Okuhata} T. Okuhata, T. Nagai, H. Taniguchi, K. Satoh, T. Itotl, Y. Shimizu, K. Miyagawa, Y. Ishii, N. Tajima, and R. Kato, High-pressure study of a doped-type organic superconductor, $\kappa$-(BEDT-TTF)$_{4}$Hg$_{2.89}$Br$_{8}$,
\href{https://link.springer.com/article/10.1007/BF02679566}{J. Low Temp. Phys.‌ {\bf142}, 547 (2006)}.
\bibitem{55Wakamatsu} K. Wakamatsu, Y. Suzuki, T. Fujii, K. Miyagawa, H. Taniguchi, and K. Kanoda, Thermoelectric signature of quantum critical phase in a doped spin-liquid candidate, 
\href{https://www.nature.com/articles/s41467-023-39217-7}{Nat. Commun. {\bf14}, 3679 (2023)}.
\bibitem{56Ji2023}	H. Ji, Y. Liu, Y. Li, X. Ding, Z. Xie, C. Ji, S. Qi, X. Gao, M. Xu, P. Gao, L. Qiao, Y.-f. Yang, G.-M. Zhang, and J. Wang, Rotational symmetry breaking in superconducting nickelate Nd$_{0.8}$Sr$_{0.2}$NiO$_{2}$ films, 
\href{https://www.nature.com/articles/s41467-023-42988-8}{Nat. Commun. {\bf14}, 7155 (2023)}.
\bibitem{57Gu2020}	Q. Gu, Y. Li, S. Wan, H. Li, W. Guo, H. Yang, Q. Li, X. Zhu, X. Pan, Y. Nie, and H.-H. Wen, Single particle tunneling spectrum of superconducting Nd$_{1-x}$Sr$_{x}$NiO$_{2}$ thin films, 
\href{https://www.nature.com/articles/s41467-020-19908-1}{Nat. Commun. {\bf11}, 6027 (2020)}.
\bibitem{58Cheng2024}	B. Cheng, D. Cheng, K. Lee, L. Luo, Z. Chen, Y. Lee, B. Y. Wang, M. Mootz, I. E. Perakis, Z.-X. Shen, H. Y. Hwang, and J. Wang, Evidence for $d$-wave superconductivity of infinite-layer nickelates from low-energy electrodynamics, 
\href{https://www.nature.com/articles/s41563-023-01766-z}{Nat. Mater. {\bf23}, 775 (2024)}.


\end{thebibliography}
\end{document}